\documentclass{jetpl}
\twocolumn
\title{Soft topological objects in topological media}

\rtitle{Soft topological objects in topological media}

\sodtitle{Soft topological objects in topological media}

\author{Jukka I. V\"ayrynen $^{*}$\/\thanks{e-mail:jukka.vayrynen@helsinki.fi}
and G.E. Volovik $^{*+}$
\/\thanks{e-mail: volovik@boojum.hut.fi}
}

\rauthor{Jukka I. V\"ayrynen, G.E.Volovik}

\sodauthor{V\"ayrynen, Volovik}

\address{ 
$^{*}$ Low Temperature Laboratory, Aalto University School of Science, P.O. Box 15100, FI-00076 AALTO, Finland
\\
$^+$ Landau Institute for Theoretical Physics RAS, Kosygina 2,
119334 Moscow, Russia
  }

\dates{January, 2011}{*}

\abstract{
Topological invariants in terms of the Green's function in momentum and real space determine
properties of smooth textures within topological media. In space dimension d=1 the topological invariant 
$N_3$ in terms of the Green's function ${\cal G}(\omega,k_x,x)$ determines  the fermion number of the kink, while in space dimension d=3 the topological invariant $N_5$ in terms of the Green's function ${\cal G}(\omega,k_x,k_y,k_z,z)$ determines quantization of Hall conductivity in the soliton plane within the topological insulators.
}

\begin{document}

\maketitle

\section{Introduction}

In the fully gapped  topological systems in even space dimension, the Chern-Simons terms are well defined. The prefactors of these terms are represented by topological invariants in momentum-frequency space, which are expressed in terms of the Green's function and are robust to interactions (see, e.g., Refs.
\cite{IshikawaMatsuyama1986,IshikawaMatsuyama1987,Matsuyama1987,VolovikYakovenko1989,Yakovenko1989,Volovik2003} for 2+1 systems; Refs. 
\cite{Volovik2003,WangQiZhang2010} for 4+1 systems; and Refs. \cite{Golterman1993,Bergman2011} for systems in general $2n+1$ spacetime). These invariants determine quantization of the parameters of the system, such as intrinsic Hall and spin-Hall conductivity in 2+1 and 4+1 systems. For odd space dimension the situation is more complicated, especially for time reversal invariant (TRI) systems, where the suggested Chern-Simons term formally violates the time reversal symmetry, see, e.g., discussion in Ref. \cite{YaoLee2010}.
Here we show that though these terms are not well-defined for bulk topological matter, they well describe the properties of smooth textures within the topological matter.
The well-defined topological invariants in momentum-frequency space are expressed in terms of the Green's function and robust to interactions. They also determine the quantization of the parameters of the system, in a given case these are the quantum numbers of the texture, such as
the fermion number in 1+1 system and quantized Hall or spin-Hall conductivity within the smooth interface in the 3+1 systems.

The smooth textures are analogues of the topological solitons in condensed matter systems with spontaneously broken symmetry.  There are two types of topological objects in these condensed matter systems: singular topological defects (such as domain walls, quantized vortices, hedgehogs and dislocations) and continuous structures called topological solitons, skyrmions and textures.  
As distinct from the singular topological defects,  which are described by conventional homotopy groups, textures do not have singularities in the order parameter fields and are described by relative homotopy groups (RHG), see \cite{MineevVolovik1978}.  These homotopy groups $\pi_n(R,\tilde R)$ deal with two different manifolds of the order parameter: the points within  the soliton are mapped to the order parameter space $R=G/H$, while the points  outside the soliton are mapped to the subspace $\tilde R=\tilde G/\tilde H$. The latter  is  restricted due to additional interaction which becomes important at large distances and which reduces the symmetry $G$ of the physical laws to its subgoup $\tilde G$. 
In the class of planar topological objects, domain walls are singular objects described by the group $\pi_0(R)$, while the smooth textures, such as planar solitons and Bloch walls, are described by the relative homotopy group $\pi_1(R,\tilde R)$.

 Analogues of such objects exist in topological matter. The role of singularities in the order parameter is played by the Green function's singularity in momentum space, such as nodes in the energy spectrum. For example, the analogue  of a singular wall is an interface between the gapped bulk states, which contains gapless fermions, while the analog of a smooth wall is  an interface, in which the Green's function has no singularity, i.e., 
within the nonsingular object the system remains fully gapped. The electronic structure of smooth textures  is  described by the relative homotopy groups $\pi_n(R,\tilde R)$ which deal with two manifolds:  the space $R$ of  the Green's function within  the texture and its subspace $\tilde R$ outside the texture. Outside  the texture the space of the Green's function is restricted due to some symmetry in bulk, while  within the texture this symmetry is violated
or is spontaneously broken.  Examples of the bulk symmetry are time reversal symmetry, spin rotation symmetry, crystal symmetry etc. Application of the RHG to classification of topological media has been also discussed in \cite{Turner2010}.

In this paper we consider smooth interfaces which separate the bulk states of time reversal invariant insulators and fully gapped superfuid systems, their momentum-space topological invariants and the related quantum numbers. The relative homotopy group describing the fully gapped interfaces in the $d$-dimensional  topological media is $\pi_{d+2}(R,\tilde R)$.

\section{Domain walls  and solitons in 3D topological media}

Let us consider examples of the singular and continuous interfaces in the vacuum described by Hamiltonian used in relativistic QFT:
\begin{equation}
\hat{H}=\tau_3 {\mbox{\boldmath$\sigma$}} \cdot{\bf p}
+ \tau_1 m_1+ \tau_2 m_2
 \,.
\label{eq:RelativisticHamiltonian}
\end{equation}
It represents a particle with a complex mass $m=m_1+im_2$.
For real mass, $m_2= 0$, the Hamiltonian obeys time reversal symmetry and anticommutes with the matrix $\tau_2$. The vacuum of this fermionic field  is described by the 3D  topological charge \cite{Volovik2010a} 
\begin{equation}
 N^K = \frac{1}{4\pi^2} ~
{\bf tr}\left[K \int_{\rm BZ}    d^3k 
~ {\cal H}^{-1}\partial_{[k_x} {\cal H}~
{\cal H}^{-1}\partial_{k_y} {\cal H}~ {\cal H}^{-1}\partial_{k_z]}  {\cal
H}\right]\,.
\label{3DTopInvariant_tau}
\end{equation} 
Here  the momentum space integral is over $\mathbb{R}_3$ in translational invariant systems, and over the Brillouin zone in crystals; $K$ is the matrix anticommuting with ${\cal H}$, which in a given case is $K=\tau_2$;   we use antisymmetrization $f_{[ij\dots ]}=\frac{1}{n!}\sum_{S_n} (-1)^P f_{PiPj\dots }$ over $n$ indices. The invariant 
\eqref{3DTopInvariant_tau} is valid also for time  reversal invariant superfluids/superconductors, such as $^3$He-B.
For the Hamiltonian \eqref{eq:RelativisticHamiltonian} with $m_2=0$ one has $N^K=m_1/|m_1|$.
The interfaces --  domain walls and solitons -- are described by the coordinate dependent mass
 \begin{equation}
\hat{H}=\tau_3 {\mbox{\boldmath$\sigma$}} \cdot{\bf p}
+ \tau_1 m_1(z) + \tau_2 m_2(z)
 \,,
\label{eq:B-phaseBound}
\end{equation}
where $z$ is the coordinate normal to the plane of the interface.

{\it Singular walls}:
The singular interface is the domain wall separating the bulk states
 with different values of $N^K$, within which the symmetry $K$ is obeyed.  
For the Hamiltonian \eqref{eq:RelativisticHamiltonian} this means that the mass remains real throughout the interface,  $m_2(z)\equiv 0$, i.e., the time reversal symmetry is obeyed for the whole interface, while the mass $m_1$ crosses zero and changes sign within the interface, $m_1(-\infty)=-m_1(\infty)$.
Since the bulk states have different topological charges, they cannot be connected adiabatically, and thus the domain wall necessarily contains gapless fermion zero modes.  That is why such interface is considered as singular.  It is the analog of the domain wall in ferromagnets in which magnetization
crosses zero value. The singular interfaces and the gapless fermion modes inside them have been discussed for superfluid $^3$He-B in \cite{SalomaaVolovik1988,Volovik2009a}. On the topological and non-topological kinks and domain walls in Grand Unified Theories (GUT), see the book by Vachaspati \cite{Vachaspati2006}. 

{\it Solitons and nonsingular walls}:
The nonsingular interface is obtained when the mass becomes complex  within the interface, $m_2(z)\neq 0$.  The time reversal symmetry is violated within the smooth interface and the spectrum becomes fully gapped everywhere. The interface where  the phase of $m_1+im_2$ changes by $\pi$ is the analog of the Bloch and Neel domain walls in ferromagnets, where the magnetization is nowhere zero and the orientation of the magnetization continuously changes across the wall. It is also the analogue of the topological soliton in $^3$He-A and $^3$He-B. 
Contrary to the singular interfaces where the symmetry is restored in the core, in continuos interfaces the symmetry is smaller than that outside the interface.  In relativistic theories, such interfaces have been considered by Wilczek \cite{Wilczek1987} and also discussed in the book \cite{Vachaspati2006}. Wilczek paid attention to the difference between the singular configuration which has gapless modes, and continuous configuration, which is fully gapped. For the gapless vacua in GUT this means that there are more massless particles outside such an interface than inside it \cite{Vachaspati2006}.

 \section{ 5-form invariant for smooth interface in 3D topological media}

 In odd space dimension the continuous texture can be described by the topological invariant which characterizes the group $\pi_{d+2}(R,\tilde R)$. For space dimension $d=3$
one has the group $\pi_5(R,\tilde R)$ with the invariant
\begin{equation}
\begin{split}
N_5 =\frac{1}{4\pi^3 i}
&{\bf tr}\int_{\rm BZ}   d^3k\int_{-\infty}^\infty dz \int_{-\infty}^\infty d\omega 
\\
& {\cal G}\partial_{[k_x} {\cal G}^{-1}
{\cal G}\partial_{k_y} {\cal G}^{-1} 
{\cal G}\partial_{k_z}  {\cal G}^{-1}
{\cal G}\partial_{\omega}  {\cal G}^{-1}
{\cal G}\partial_{z]}  {\cal G}^{-1} \,,
\end{split}
\label{pi5}
\end{equation} 
where $z$ is coordinate across the interface. This invariant in terms of the quasiclassical matrix Green's function ${\cal G}({\bf k},z,\omega)$ is applicable  to different $d=3$ systems such as the 3D  topological insulators and superfluid $^3$He-B.
The 5-form integrals for topological media in terms of Green's functions have been discussed in 
Refs. \cite{Volovik2003,SilaevVolovik2010,WangQiZhang2010}. 
The combined momentum space and real space topology has been applied for description of  the singular topological defects and interfaces within the topological media and fermion zero modes in these objects, see
\cite{VolovikMineev1982,GrinevichVolovik1988,SalomaaVolovik1988,Volovik2003,TeoKane2010,SilaevVolovik2010,Gurarie2010},
and now we discuss this  for continuous textures.
For general continuous textures, $N_5$ may take any value, but it becomes integer or half-integer for appropriate boundary conditions when the bulk states at $z=\infty$ and $z=-\infty$ coincide, or are connected by symmetry.
 
As an example, consider Hamiltonian \eqref{eq:B-phaseBound} with $m_1(z)=M_1\cos\phi(z)$ and $m_2(z)=M_2\sin\phi(z)$, where $\phi(z)$ changes from $0$ to $\pi n$ across the interface. The gap is finite everywhere, and thus the integral \eqref{pi5} is well-defined and equals  $N_5=n/2 ~{\rm sign} (M_1M_2)$. For $n=2k$, the states are the same on two sides of the soliton and $N_5$ is an integer. For $n=(2k+1)$, the bulk states have opposite sign of mass $m_1$ and the invariant $N_5$ is a half integer.  If time reversal symmetry is not obeyed in the bulk, the integral may take arbitrary values.
 
It is important that the invariant $N_5$ is determined both by the states of bulk topological matter outside the texture and by the internal structure of the texture. This invariant is expressed in terms of the Green's function, 
and thus is robust to perturbations such as interactions. In interacting systems the single-particle Hamiltonian, such as that which enters \eqref{3DTopInvariant_tau},  is the secondary object. It is the effective Hamiltonian which belongs to the same topological class as the original interacting system, and thus can be adiabatically obtained from the interacting system. For example, one can consider the inverse Green's function at zero frequency as effective Hamiltonian, ${\cal H}_{\rm eff}({\bf k})={\cal G}^{-1}(\omega=0,{\bf k})$.

 The invariant $N_5$ can be also applied to 3D topological insulators. Let us consider the model Hamiltonian for a TRI insulator (see e.g. \cite{RosenbergFranz2010}):
\begin{equation}
{\cal H}=\gamma_\mu n_\mu({\bf k}) ~, ~\gamma_0=\tau_1~,~\gamma_i =\sigma_i\tau_3~,
 \label{H}
\end{equation}
where $n_\mu$ is a 4-vector and in a relativistic theory the matrices must be multiplied by $\tau_1$ to get the conventional $\gamma$-matrices. The particular 4-vector discussed in \cite{RosenbergFranz2010}  is:
\begin{equation}
{\cal H}=-\lambda \tau_3(\sigma_x \sin k_x+\sigma_y \sin k_y+\sigma_z \sin k_z)+ \tau_1 m_1({\bf k})~, \label{H_example}
\end{equation}
where $m_1({\bf k})=M_1- t(\cos k_x +\cos k_y  +\cos k_z)$.
Inside the texture one has
\begin{equation}
\begin{split}
&{\cal H}_{\rm texture}=-\lambda \tau_3(\sigma_x \sin k_x+\sigma_y \sin k_y+\sigma_z \sin k_z)
\\
&+ \tau_1 m_1({\bf k},z) + \tau_2 m_2({\bf k},z)~.
\end{split}
 \label{soliton}
\end{equation}
One may choose, for example, the following texture: $m_1({\bf k},z) =M_1 \cos \phi(z)  - t(\cos k_x +\cos k_y  +\cos k_z)$  and $m_2({\bf k},z) =M_2\sin \phi(z)$, with $\phi$ changing from $0$ to $\pi n$ across the interface.
For $2t<|M_1|<3t$ and large enough $|M_2|$, one obtains $N_5=n/2 ~{\rm sign} (M_1M_2)$.

\section{5-form invariant, $\theta$-term and QHE}

 For the 3D insulators the effects similar to those in axion QED take place. A $\theta$-term in the electromagnetic action has been proposed  for the time reversal invariant (TRI) insulators, see, e.g.,  \cite{Qi2008,Essin2009,RosenbergFranz2010}:
\begin{equation}
S=\frac{ e^2}{32\pi^2} \varepsilon^{\alpha\beta\mu\nu}\int d^4x~\theta F_{\alpha\beta}F_{\mu\nu}=\frac{e^2}{4\pi^2} \int d^4x~\theta {\bf E}\cdot {\bf B}~.
 \label{theta_term}
\end{equation}
In bulk insulators,  $\theta$ is space-time constant, and this term does not make sense, since the action becomes a total derivative. Moreover, the $\theta$-term violates time reversal invariance and its application to TRI systems is tricky, though in a periodic space-time the situation is clearer \cite{ChenLee2010}. However, all these problems vanish when we discuss the properties of a smooth texture within which the  time reversal invariance is violated. While the parameter  $\theta$ itself is ill-defined, the Hall conductivity in the plane of the interface is a well defined quantity, though formally according to \eqref{theta_term} it can be related to the change of $\theta$ across the interface
(see \cite{HasanMoore2010}):
 \begin{equation}
\frac {\sigma_{xy}}{\sigma_{\rm H}}=\frac{\theta(+\infty)-\theta(-\infty)}{2\pi}~~,~~\sigma_{\rm H}=\frac{e^2}{h} \,,
  \label{HallConductivity}
\end{equation} 
Using the gradient expansion of the action, the Hall conductivity in the interface is expressed in terms of the invariant $N_5$:
 \begin{equation}
\frac {\sigma_{xy}}{\sigma_{\rm H}}=N_5
 \,.
  \label{QuantizationHallconductivity}
\end{equation} 
Applying this to the Hamiltonian \eqref{eq:B-phaseBound}, where $\theta$ is related to the complex mass,   $m_2/m_1= \tan \theta$ \cite{Wilczek1987}, for a texture where $m_2$ changes from $-m_0$ to $m_0$, 
one obtains
\begin{equation}
\begin{split}
\frac {\sigma_{xy}}{\sigma_{\rm H}}=&\frac{ 1}{4\pi^3 i}
{\bf tr}\int_{\rm BZ}   d^3k\int_{-m_0}^{m_0} dm_2 \int_{-\infty}^\infty d\omega 
\\
& {\cal G}\partial_{[k_x} {\cal G}^{-1}
{\cal G}\partial_{k_y} {\cal G}^{-1} 
{\cal G}\partial_{k_z}  {\cal G}^{-1}
{\cal G}\partial_{\omega}  {\cal G}^{-1}
{\cal G}\partial_{m_2]}  {\cal G}^{-1}.
\end{split}
\label{example}
\end{equation}
This transforms to the integer-valued $N_5$ in the limit of large $m_0$.
The same is applied  for the more general Hamiltonian ${\cal H}=\gamma_\mu n_\mu({\bf k}) +\gamma_5 m_2$, $\gamma_5 = \tau_2$. 

According to \eqref{QuantizationHallconductivity}, the Hall conductivity is determined both by the properties of bulk states outside the interface and by the internal structure of the interface.
For the texture inside a TRI topological insulator, i.e., in the system which is TRI at $z=\pm \infty$, the Hall conductivity is quantized.
Note that in this system the integral $N_5$ 
is a topological  invariant, which belongs to the group $\mathbb{Z}$, i.e., it can take any integer or half-integer value, as distinct from the $\mathbb{Z}_2$ nature  of the bulk insulator, where the parameter $\theta$ is ill defined. In other words, the group $\pi_5=\mathbb{Z}$ with its invariant $N_5$ describes the topology of the solitons within the topological insulator, rather than the insulators themselves.

\section{1D skyrmions and their topological and quantum numbers}

The relative homotopy group $\pi_3(R,\tilde R)$ with the 3-form integral $N_3$ in terms of the Green's function  ${\cal G}(\omega,k_x,x)$,
\begin{equation}
\begin{split}
N_3 =\frac{1}{4\pi^2} 
{\bf tr}\int_{-\infty}^\infty dx \int_{-\infty}^\infty d\omega \int_{\rm BZ}    dk_x
\\
{\cal G}\partial_{[k_x} {\cal G}^{-1}
{\cal G}\partial_{\omega}  {\cal G}^{-1}
{\cal G}\partial_{x]}  {\cal G}^{-1} \,,
\end{split}
\label{pi3}
\end{equation} 
describes 1D skyrmions in 1D gapped topological systems. 
These solitons have  quantum numbers such as fermionic charge and quantized electric charge \cite{JackiwRebbi1976,Su1979,GoldstoneWilczek1981}. In general these charges are expressed in terms of the topological charge $N_3$. Let us now consider the electric charge.  Typically this charge is related  to 1D $\theta$-term in the action,
\begin{equation}
S=\frac{1}{2\pi}\int dxdt \theta \varepsilon^{\alpha\beta}\partial_{\alpha}A_\beta \,.
\label{1+1}
\end{equation} 
Again, this $\theta$-term does not make much sense for constant $\theta$, but it becomes meaningful  for the inhomogeneous order parameter within the soliton where the time reversal symmetry is violated. Assuming $A_0$ is constant in space, 
one obtains the solitonic electric charge:
\begin{equation}
S=\frac{1}{2\pi}\int dxdt \partial_x \theta  A_0 =\frac{1}{2\pi}\int dt (\theta(+\infty)-\theta(-\infty))  A_0 \,,
\label{1+1b}
\end{equation}  
 \begin{equation}
q= \frac{\theta(+\infty)-\theta(-\infty)}{2\pi} 
 \,.
  \label{ChargeQuantization}
\end{equation} 
As distinct from the ill-defined $\theta$, the fermionic charge of the smooth structure is well defined. It is expressed in terms of the Green's function, and -- using the gradient expansion -- in terms of the quasiclassical Green's function ${\cal G}(\omega,k_x,x)$:
\begin{equation}
\begin{split}
q=&\frac{1}{4\pi^2}{\bf tr}\int  dx  \int_{-\infty}^\infty d\omega \int_{\rm BZ}    dk_x~
\\
& {\cal G}\partial_{[k_x} {\cal G}^{-1} {\cal G}\partial_{\omega}  {\cal G}^{-1} {\cal G}\partial_{x]}  {\cal G}^{-1}, 
   \label{ChargeQuantization2}
   \end{split}
\end{equation} 
which is the invariant $N_3$. Thus one obtains the general relation between the Green's function topological invariant $N_3$ characterizing  the smooth structure and its fermionic charge $q$:
 \begin{equation}
q= N_3
 \,.
   \label{ChargeQuantization3}
\end{equation} 
Eq. \eqref{ChargeQuantization3} is analogous to equation \eqref{QuantizationHallconductivity} for quantization of Hall conductivity
in the smooth interface within the 3D TRI topological insulators. It is valid for interacting systems as well.

Consider as an example the soliton in the following 1D system \cite{Qi2008}  
\begin{equation}
\hat{H}=\tau_3 p_x
+ \tau_1 m_1  + \tau_2 m_2 
 \,,
\label{eq:1D}
\end{equation}
which is time reversal invariant for $m_2=0$.
For the soliton  with $m_{1}(x)= M\cos \varphi(x)$,  
$m_{2}(x)=M\sin\varphi(x)$, where $\varphi$ is going from 0 to $\pi n$, one obtains $N_3=n/2$ and  thus the charge  of this soliton must be $q=n/2$. The fermion number of the domain wall with $m_2=0$, $m_1(+\infty)=-m_1(-\infty)$ has been discussed by Jackiw and Rebby \cite{JackiwRebbi1976}, who  got the fermionic number $1/2$, i.e. $q=\pm 1/2$. This agrees with the topological charge $N_3=\pm 1/2$ of the soliton obtained by softening of the domain wall when the imaginary mass is added. Such softening does not change the boundary conditions at infinity and thus the fermionic charge may change only by integer number -- the number of fermions.

\section{Discussion}

We found a connection between topological invariants describing the smooth textures inside the topological media and their quantum numbers. We considered only one type of textures in (1+1)D and (3+1)D media.
The other types of the continuous topological objects in topological matter are also possible. 
 The relative homotopy group $\pi_{5}(R,\tilde R)$  with the 5-form integral $N_5$ in \eqref{pi3} in terms of the Green's function ${\cal G}(\omega,k_x,k_y,x,y)$  
\begin{equation} 
 \begin{split}
N_5 =\frac{1}{4\pi^3 i}
&{\bf tr}\int_{\rm BZ}   d^2k \int d^2x \int d\omega 
\\
& {\cal G}\partial_{[k_x} {\cal G}^{-1}
{\cal G}\partial_{k_y} {\cal G}^{-1} 
{\cal G}\partial_{\omega}  {\cal G}^{-1}
{\cal G}\partial_{x}  {\cal G}^{-1} 
{\cal G}\partial_{y]}  {\cal G}^{-1}\,,
\end{split}
\label{2Dskyrmion}
\end{equation} 
describes the 2D skyrmions in 2D gapped topological systems, such as $^3$He-A and planar phase. The quantum numbers of 2D skyrmions and the corresponding Chern-Simons terms in the action have been considered in \cite{VolovikYakovenko1989,Yakovenko1989}.
 The relative homotopy group  $\pi_7(R,\tilde R)$ with the 7-form integral $N_7$ in terms of the Green's function ${\cal G}(\omega,k_x,k_y,k_z,x,y,z)$ describes the 3D skyrmions in the 3D gapped topological systems.  

The mixed real-space and momentum-space topology can be applied for skyrmions and solitons in relativistic quantum field theories such as GUT, QCD, electroweak theory and theory of chiral and color quark matter. In particular, the $\theta$-term and axion electrodynamics \cite{PecceiQuinn1977} can be treated in the same manner as for $^3$He-B and TRI insulators, using integrals \eqref{pi5} and \eqref{3DTopInvariant_tau}. The fermionic charges of skyrmions and other textures are related to  the topological invariants expressed in terms of the fermionic propagator.

 This work is supported in part by the ERC (Grant No. 240362-Heattronics) and the Academy of Finland, Centers of excellence program 2006--2011.

\section{Appendix}

In this appendix, we show how Eq. \eqref{QuantizationHallconductivity}
is obtained through the gradient expansion. By integrating out the
fermions in the path integral, we obtain a current
\begin{equation}
j^{\gamma}=\frac{\delta}{\delta A_{\gamma}}i\mathbf{Tr}\ln G=ie\mathbf{tr}\int\dfrac{d^{3}kd\omega}{(2\pi)^{4}}G^{-1}\partial_{k_{\gamma}}G \,,\label{eq:currentgeneric}\end{equation}
which, at low energies, can be expanded in powers of gradients. Using
Wigner transformed (or quasiclassical) Green functions $\mathcal{G}$
and the Moyal product rule, we obtain a gradient series for the Wigner
transformed $G^{-1}$. From this series we will extract the part contributing
to the current\[
j^{\gamma}=\dfrac{e^{2}}{8\pi^{2}}\varepsilon^{\alpha\beta\gamma\delta}F_{\alpha\beta}\partial_{\delta}\theta\,.\]
This is second order in derivatives and we obtain
\begin{flalign*}
j^{\gamma}= & \dfrac{e^{2}}{2i}F_{\alpha\beta}{\bf tr}\int \dfrac{d^{3}kd\omega}{(2\pi)^{4}}  \partial_{k_{\gamma}}\mathcal{G}
 \left\{ \partial_{[\delta}\mathcal{G}^{-1}
\partial_{k_{\delta}]}\partial_{k_{\alpha}}\left(\mathcal{G}\partial_{k_{\beta}}\mathcal{G}^{-1}\right) \right. \\
 & -\mathcal{G}^{-1}\partial_{k_{\alpha}}\mathcal{G}\partial_{k_{\beta}}\mathcal{G}^{-1}\partial_{[\delta}\mathcal{G}\partial_{k_{\delta}]}\mathcal{G}^{-1}\\
 & -\mathcal{G}^{-1}\partial_{[\delta}\mathcal{G}\partial_{k_{\delta}]}\partial_{k_{\alpha}}\mathcal{G}\partial_{k_{\beta}}\mathcal{G}^{-1}\\
 & -\partial_{k_{\beta}}\mathcal{G}^{-1}\partial_{k_{\alpha}}\left(\partial_{[\delta}\mathcal{G}\partial_{k_{\delta}]}\mathcal{G}^{-1}\right)\\
 & \left.-\frac{1}{2}\mathcal{G}^{-1}\partial_{k_{\beta}}\partial_{[\delta}\mathcal{G}\partial_{k_{\delta}]}\partial_{k_{\alpha}}\mathcal{G}^{-1}\right\}\,,\end{flalign*}
where $\mathcal{G}$ is evaluated in zero external field. For a linear
Hamiltonian we have $\partial_{k_{\alpha}}\partial_{k_{\beta}}\mathcal{G}^{-1}=0$
and finally obtain the action \begin{flalign*}
S= & \frac{e^{2}}{16\pi^{2}}\int d^{3}xdt\varepsilon^{\alpha\beta\gamma\delta}F_{\alpha\beta}A_{\gamma}\partial_{\delta}\theta,\\
\partial_{\delta}\theta= & \frac{1}{2\pi^{2}i}\mathbf{tr} \int_{\rm BZ} d^{3}k \int_{-\infty}^\infty d\omega\\
 & \mathcal{G}\partial_{[\delta}\mathcal{G}^{-1}\mathcal{G}\partial_{k_{\lambda}}\mathcal{G}^{-1}\mathcal{G}\partial_{k_{\mu}}\mathcal{G}^{-1}
\mathcal{G}\partial_{k_{\nu}}\mathcal{G}^{-1}\mathcal{G}\partial_{k_{\rho}]}\mathcal{G}^{-1}\,,\end{flalign*}
which leads to Eq. \eqref{QuantizationHallconductivity}.

In exactly the same fashion, we obtain Eq. \eqref{ChargeQuantization3}.
In this case the current of Eq. \eqref{eq:currentgeneric} is obtained
from the first order gradient expansion and reads
\begin{equation}
\begin{split}
j^{\gamma}=\dfrac{e}{4\pi^2}\mathbf{tr} \int_{-\infty}^\infty d\omega \int_{\rm BZ}    dk_x  \mathcal{G}^{-1}
\left(\partial_{[\alpha}\mathcal{G}\partial_{k_{\alpha]}}\mathcal{G}^{-1}
\right)
\partial_{k_{\gamma}}\mathcal{G}\,.   \end{split}
\end{equation}

\end{document}